\documentclass{elsart}
\usepackage[dvips]{graphicx}

\begin{document}
\begin{frontmatter}
\title{Chaotic dynamics and superdiffusion in a Hamiltonian system
with many degrees of freedom}

\author{V. Latora\thanksref{vito}}
and
\author{A. Rapisarda\thanksref{andrea}}

\address{Dipartimento di Fisica Universit\'a di Catania,
Corso Italia 57 I 95129 Catania, Italy}

\author{S. Ruffo\thanksref{stefano}}
\address{
Laboratoire de Physique, Ecole Normale Sup\'erieure de Lyon\\
46 all\'ee d'Italie 69364 Lyon Cedex 07 France, INFM and
INFN Firenze, Italy
}

\thanks[vito]{E-mail: latora@ct.infn.it}
\thanks[andrea]{E-mail: andrea.rapisarda@ct.infn.it}
\thanks[stefano]{
{On leave from Dipartimento di Energetica ``S. Stecco", 
Universit\'a di Firenze, Via S. Marta, 
3 I-50139, Firenze, Italy} 
, E-mail:ruffo@avanzi.de.unifi.it}

\begin{abstract}
We discuss recent results obtained for the Hamiltonian Mean Field model.
The model describes a system of N fully-coupled particles in one dimension and
shows a second-order phase transition from a clustered phase
to a homogeneous one when the energy is increased. 
Strong chaos is found in correspondence to the critical point on top 
of a weak chaotic regime which characterizes the motion at low energies.  
For a small region around the critical point, we find anomalous 
(enhanced) diffusion and L\'evy walks in a transient temporal 
regime before the system relaxes to equilibrium.  

\end{abstract}

\end{frontmatter}

\section{Introduction}

In this contribution we present a novel study of chaos and 
anomalous diffusion in the dynamics 
of the single-particle motion in the Hamiltonian Mean Field (HMF) model.
This model describes N classical particles moving on the unit circle 
and interacting through an infinite range potential. It 
has been recently studied both analitycally 
and numerically~\cite{ruffo,latora98,latora99,prl99}. 
At low energy the particles are clustered and, as the energy 
is increased, the system undergoes
a second-order phase transition, reaching a high temperature ``gaseous"
phase where the particles are homogeneously distributed on the circle.
The motion is chaotic and the Lyapunov exponent reaches a maximum at the
critical point. 

Particle motion is superdiffusive for a range of energies 
close to the critical point, but only in a transient quasi-stationary 
regime preceeding equilibration.  Particles can evaporate from the cluster
showing L\'evy walks. Close to the critical point, if the continuum 
($N\rightarrow \infty$) limit is performed before the $t \to \infty$ 
limit, canonical equilibrium is never reached and quasi-stationary
superdiffusive states live forever~\cite{ruffo,latora98,latora99,prl99}.
Though anomalous diffusion and L\'evy walks have been widely 
studied in the recent literature in models with a few degrees 
of freedom~\cite{levypro,kla}, there are very few studies 
for many-degrees-of-freedom 
systems~\cite{kaneko,torc}. The HMF model
has the advantage of having an exact solution in the canonical ensemble, and 
therefore microscopic dynamics can be put in connection to macroscopic 
features, thus offering a very interesting perspective~\cite{latora99,prl99}.
  
In the following sections we remind the reader the details
of the model and then we discuss superdiffusion in 
connection to microscopic chaotic dynamics.

\section{The model}

HMF describes a system of $N$ classical particles moving on a circle,
characterized by the angles $\theta_i$ and the conjugate momenta 
$p_i$.
Each particle interact with all the others according to the following 
Hamiltonian:  
\begin{equation}
        H(\theta,p)=K+V ~,
\end{equation}
where 
\begin{equation}
       K= \sum_{i=1}^N  {{p_i}^2 \over 2} ~~~~~~~ 
       V= {1\over{2N}} \sum_{i,j=1}^N  [1-cos(\theta_i -\theta_j)]
\end{equation}

are the kinetic and potential energy. 
If we define a spin vector associated to each particle
${\bf m}_i=[cos(\theta_i), sin(\theta_i)]$ 
and a total magnetization ${\bf M}={\frac{1}{N}}\sum_{i=1}^N {\bf m}_i$
the Hamiltonian then describes $N$ classical fully coupled spins,  
similarly to the XY model.  
With this interpretation, the system is a ferromagnet at low
energy~~ and shows a second-order phase transition ~~at the critical 
energy~~ $U_c=E_c/N=0.75$ (corresponding to the critical
temperature $T_c=0.5$ in the canonical 
ensemble)~\cite{ruffo,latora98,latora99}).

\begin{figure}
\includegraphics[height=0.4\textheight,angle=-90]{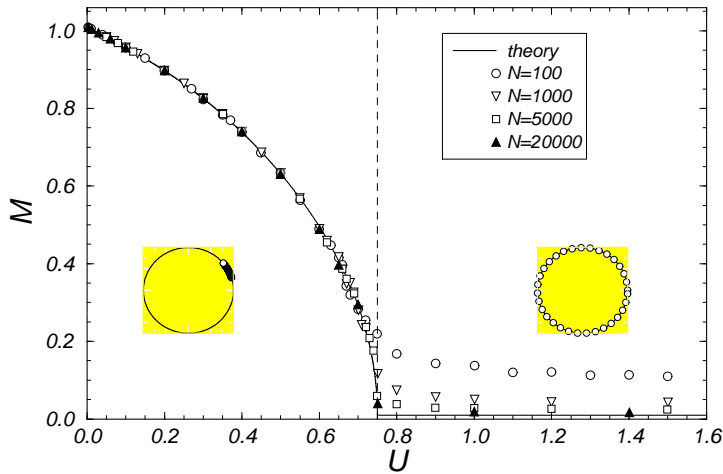}
\caption{Total magnetization $M$  as a function of the energy per particle $U=E/N$. 
Numerical simulations are compared with the canonical prediction (full curve). The insets
show a schematic pictorial view of the condensed and the homogeneous  phase. See text.}
\label{fig.1}
\end{figure}

We show in fig.1 the total magnetization vs. energy $U=E/N$. Numerical
simulations (dots) are compared with the canonical ensemble result (full line).
Pictorial views of the low energy ferromagnetic (clustered)
phase and high energy paramagnetic (gaseous) phases are shown in the insets. 
The equations of motion for the $N$ particles are 
\begin{equation} 
\dot{\theta_i}={p_i}, ~~~\dot{p_i}  = - M sin(\theta_i - \phi ) ~~~,~~~ 
i=1,...,N~~~,
\label{eqmoto} 
\end{equation}
where $(M,\phi)$ are respectively the modulus and the phase 
of the total magnetization vector $\bf M $. 
These equations are formally equivalent to those of
a perturbed pendulum.                                   
To study relaxation to canonical equilibrium, we solve these equations 
on the computer using fourth order symplectic algorithms 
(the details can be found in Refs~\cite{latora98,latora99}).  
We start the system in a given initial state 
and we compute ${\theta_i, p_i}$ at each time step, and from them   
the total magnetization $M$ and temperature $T$
(through the relation $T=2<K>/N$).  
We consider systems with an increasing size $N$ and different 
energies $U=E/N$.
The single-particle motion is given by Eq.(~\ref{eqmoto}), if we
consider the mean-field pair $(M,\phi)$ as an {\it external} time
dependent forcing.
In the next section we focus our attention on the dynamics of the 
single-particle motion and we show that it is chaotic and 
superdiffusive.

\section{Chaotic motion and superdiffusion}

In this section we discuss the numerical results. 
In fig.~2 we show single-particle phase spaces for three different 
energies: below (2a), close to
 (2b) and above (2c) the critical energy $U_c=0.75$. 
While for $U=0.1$ the particle never escapes the cluster and moves 
on the chaotic web near the center of the nonlinear resonance, close to 
the critical point (at $U_c=0.69$) it escapes frequently from the
cluster and is then trapped again, exhibiting the well known
phenomenon of {\it separatrix crossing}. Above the critical energy one has 
mainly a ballistic regime, separatrix crossings are rarely observed and
particles interact very weakly.

\begin{figure}
\includegraphics[height=0.58\textheight,angle=-90]{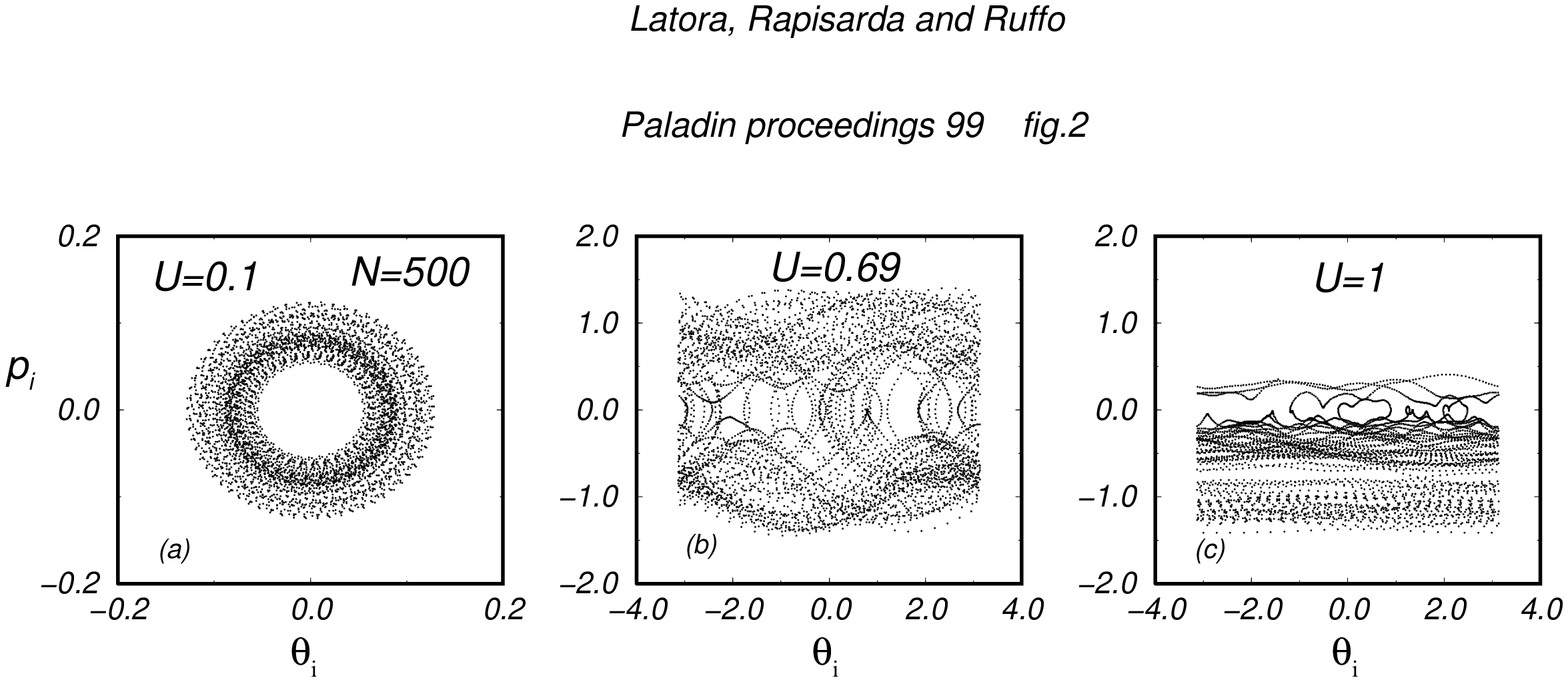}
\caption{Single-particle phase space for three different  energy regimes:
$U=0.1$ ($<U_c$), $U=0.69$   (close to $ U_c$)  and $U=1$  ($>U_c$). See text for 
more details.}
\end{figure}

We show in fig.~3 the largest Lyapunov exponent $\lambda_{sp}$
for the single-particle motion as a function of the total energy 
per particle. The motion is chaotic and has a peak at the critical 
point, but $\lambda_{sp}$ is much smaller than the largest Lyapunov exponent 
calculated for the full dynamics in Refs.~\cite{latora98,latora99}. 
The reason is that in the calculation of the Lyapunov of the full
system one gets the additional contribution to the instability of the off-diagonal
terms of the Jacobian. We show in the inset a log-log plot with a fit 
at energies less than $0.3$,  which is a power law, i.e. $\lambda\sim U^{1/3}$; again 
quite far from the $\sqrt{U}$ law found for the full dynamics. A very 
weak $N$ dependence is present is this region. In the high energy region
$\lambda_{sp}$ vanishes with $N$, with the ``universal" law $\lambda_{sp} 
\sim N^{-1/3}$, which is found in many different contexts~\cite{gravity}. This result is
displayed in fig.3(b) for $U=2$.

\begin{figure}
\includegraphics[width=0.5\textwidth,height=0.3\textheight,angle=0]{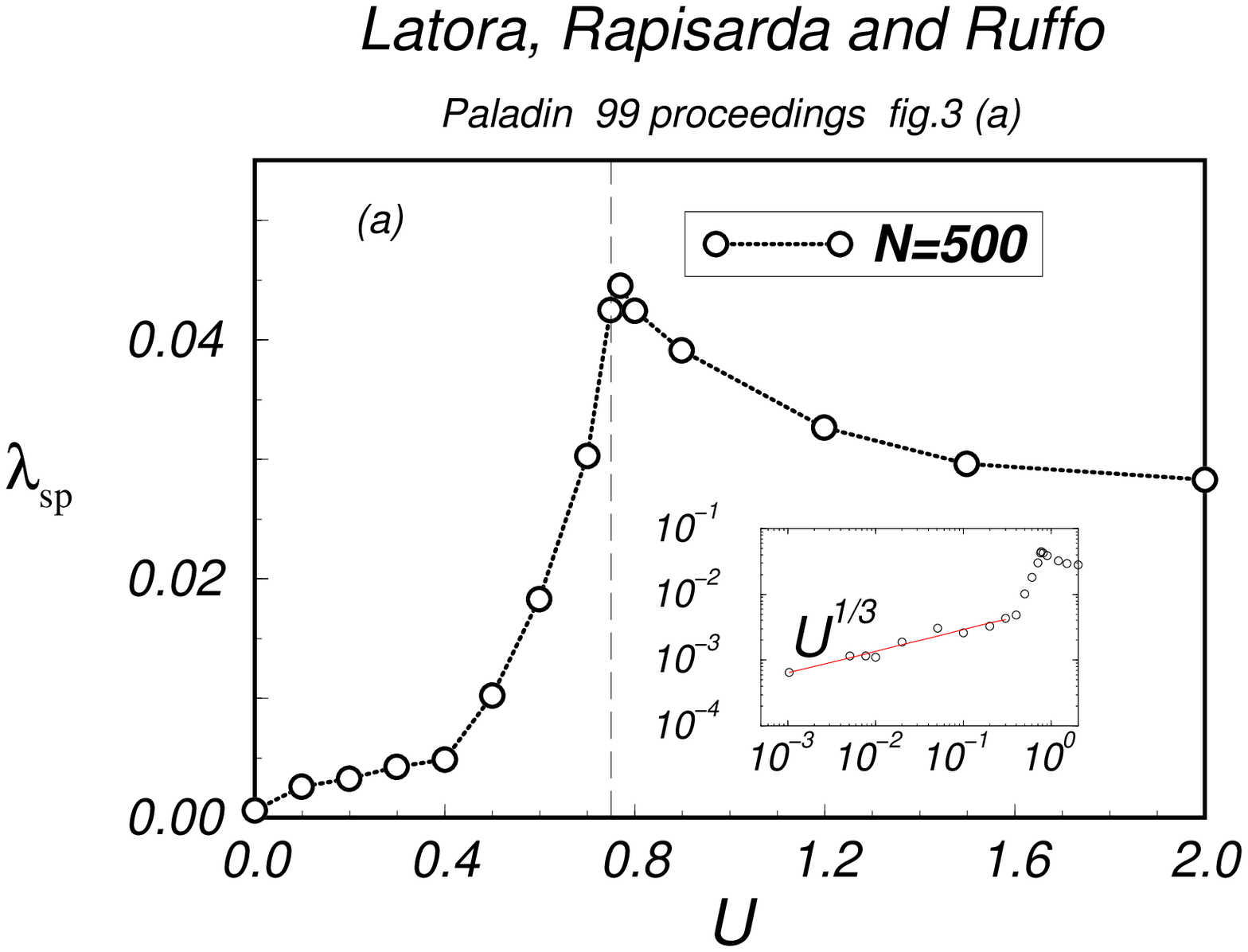}
\includegraphics[width=0.5\textwidth,height=0.3\textheight,angle=0]{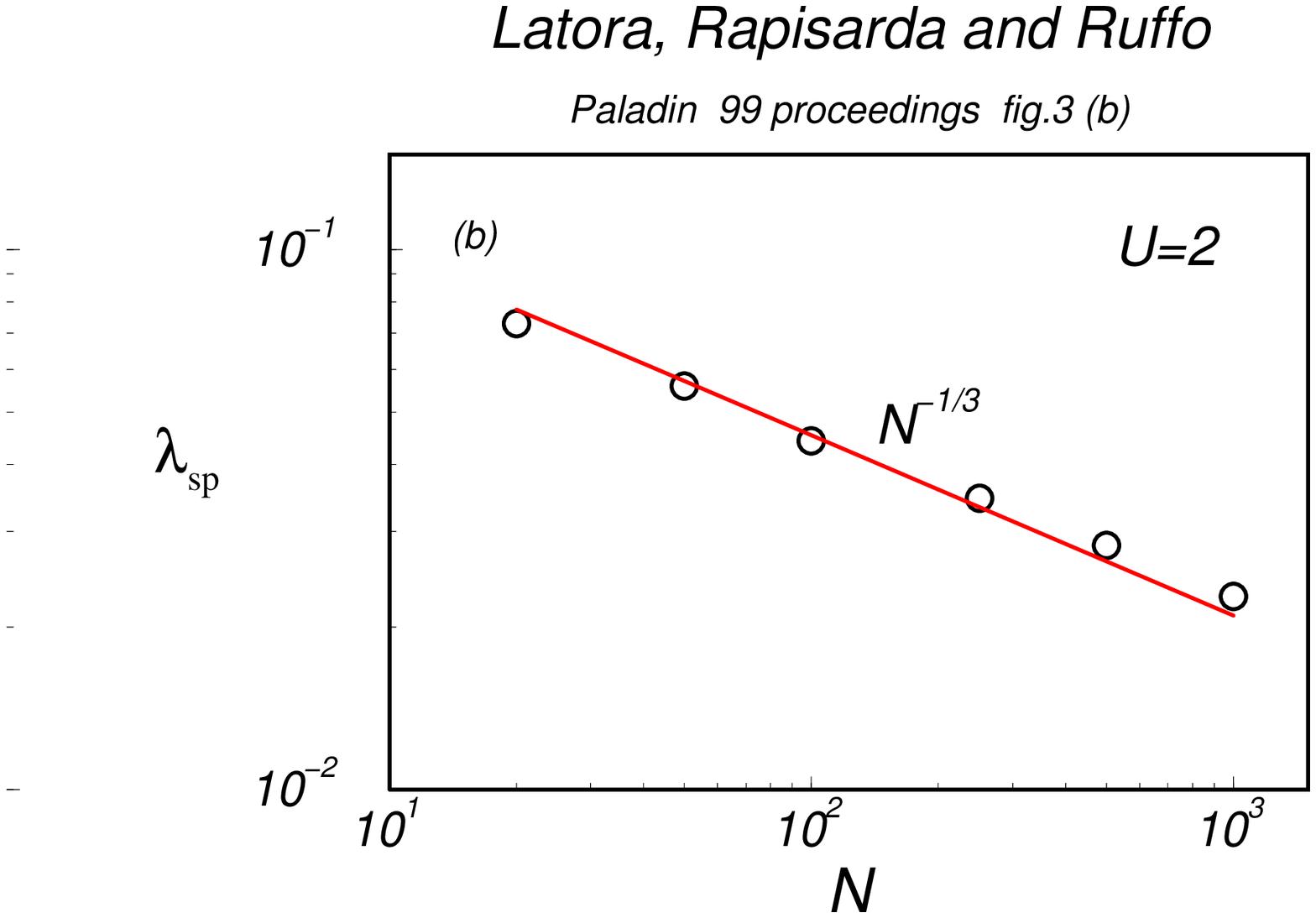}
\caption{(a) We show the largest Lyapunov exponent for the
 single-particle as a function of $U$. 
In the inset we show the behavior at small energy in a log-log plot, see text. (b) We report numerical
simulations (open circles) that show that $\lambda_{sp}$  scales like
 $N^{-1/3}$ (straight line).
}
\end{figure}

Diffusion and transport of a particle in a medium or in a fluid 
flow are characterized by the mean square displacement $\sigma^2(t)$
in the long-time limit.
In general, one has 
\begin{equation}
\label{anoma}
    \sigma^2(t) \sim  t^{\alpha}
\end{equation}
with $\alpha=1$ for normal diffusion.   
All the processes with $\alpha \ne 1$ are termed anomalous 
diffusion, namely subdiffusion for $0<\alpha<1$ and 
superdiffusion for $1<\alpha<2$. 

In order to study anomalous diffusion in HMF we 
follow the dynamics of each particle starting the system in a 
``water bag'', i.e. a far-off-equilibrium initial condition
obtained by putting all the particles at $\theta_i=0$ and
giving them a uniform distribution of momenta with a
finite width centered around zero. 
We compute the variance of the one-particle angle $\theta$ 
according to the expression
\begin{equation}
    \sigma^2(t) = < (\theta - <\theta> )^2 >
~~~~~~~~,
\end{equation}
where  $< ~.~>$  indicates  the average over the $N$ particles, 
and we fit the value of the exponent $\alpha$ in Eq. (\ref{anoma}).

In fig.~4(a) we show the motion of four particles in the same
simulation as a function of time for $N=500$ and $U=0.69$.  
The figure shows that
the erratic motion is not brownian and L\'evy walks are present. 
Each particle experiences in time localized motion and free walks.
In fig.4(b) we show the mean square displacement as a function
of time. For energies close to the critical one the exponent $\alpha$
differs from 1, as it should be for normal diffusion,  and takes
values close to 1.4. One has therefore superdiffusion, which however 
turns again into normal after a crossover time. This
behavior does not seem to depend strongly on energy, although the
superdiffusion exponent $\alpha$ is slightly energy dependent.
In Ref. ~\cite{prl99} we have shown that this crossover time 
coincides with the equilibration one and that by using the model of 
Ref.~\cite{kla} it is possible to interpret this dynamics in terms 
of L\'evy walks.

\begin{figure}
\includegraphics[width=0.5\textwidth,height=0.3\textheight,angle=0]{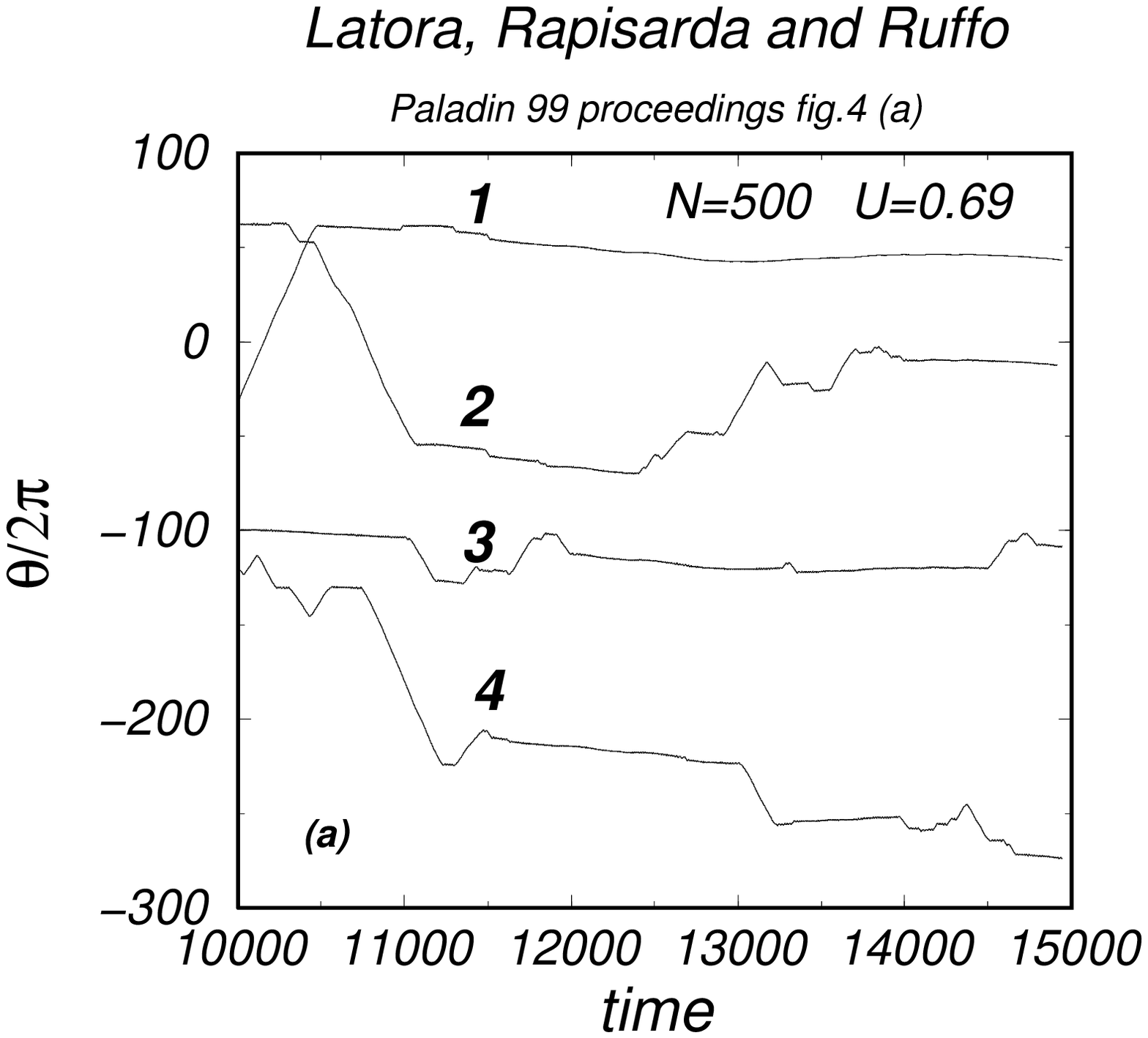}
\includegraphics[width=0.5\textwidth,height=0.3\textheight,angle=0]{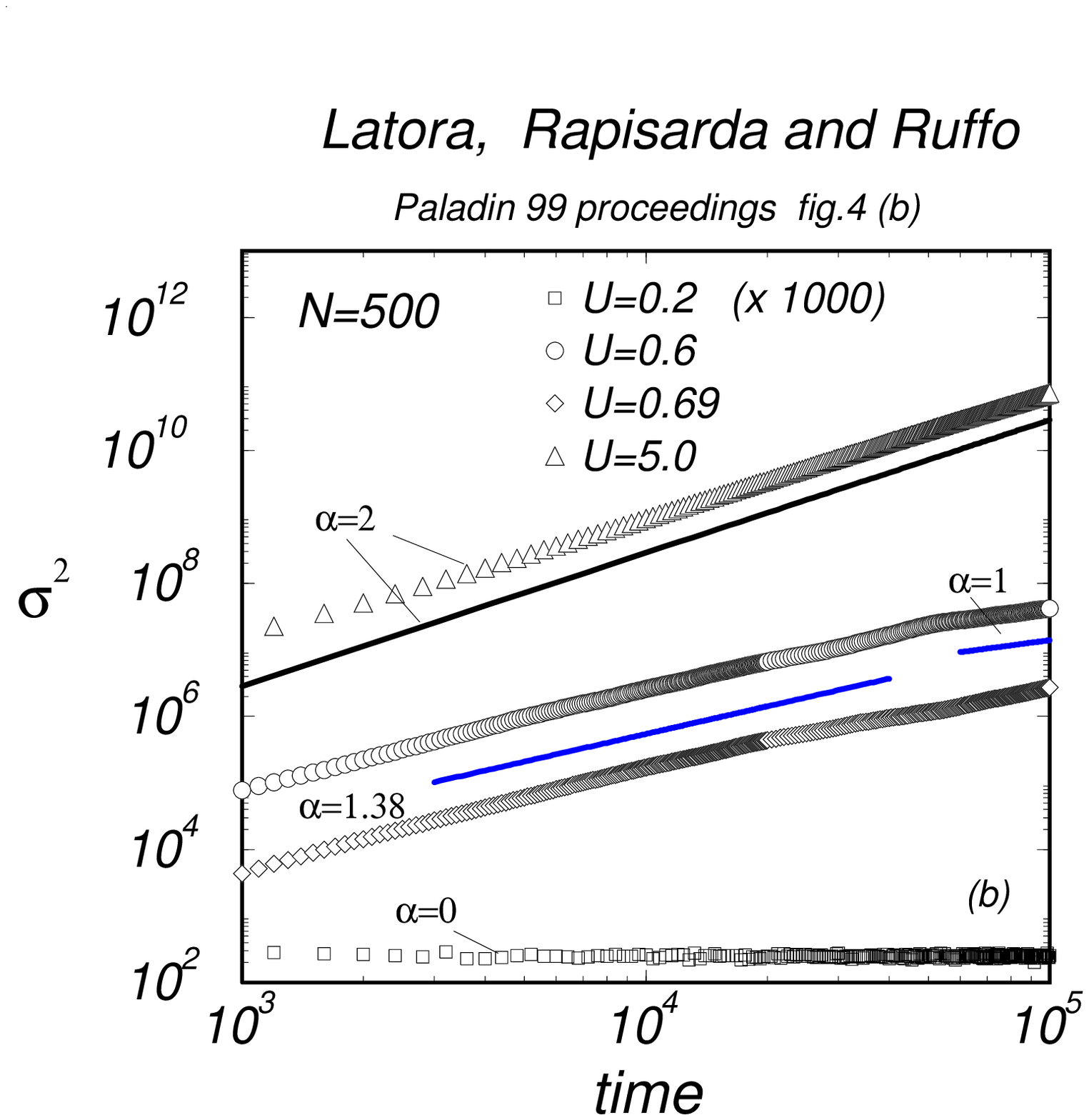}
\caption{(a) L\'evy walks for different test particles at $U=0.69$. (b) 
Variance of the single-particle displacement as a function 
of time for four different energies.
}
\end{figure}

\section{Conclusion}

We have discussed chaotic and superdiffusive motion in a 
N-body Hamiltonian system which exhibits 
a second-order phase transition.
The single-particle motion is shown to be chaotic 
for energies smaller than the critical one. 
Superdiffusion is present in a transient out-of-equilibrium regime
for energies close to the critical one, where 
the system is strongly chaotic and quasi-stationary states exist. 
The crossover time from enhanced diffusion to normal 
coincides with the equilibration time which diverges with $N$.
This is true for fully coupled systems, but should appear also if
the interaction decays with the distance among the particles with
a power law~\cite{celia}.
Our model lies in the class of self-consistent many-degrees-of-freedom
dynamics, to which also belongs the single-wave model described in 
Ref.~\cite{delca}. This is the closest contact of our results to
the L\'evy-walk behavior found in fluid experiments~\cite{solomon}.

\begin{ack}
We dedicate this work to the memory of Giovanni Paladin who was an 
enthusiastic pioneer in the study of chaos and statistical mechanics,
and the first one to explain to one of the authors (S.R.) the concept
of anomalous diffusion. We thank C. Anteneodo, M. Antoni,M. Baranger,
 E. Barkai, 
 D. del-Castillo-Negrete, P. Grigolini, 
A. Torcini and C.Tsallis for useful discussions.S.R.
 thanks ENS-Lyon for financial support.
\end{ack}

\end{document}